\documentclass[10pt,aps,prd,showpacs,nofootinbib,eqsecnum,preprintnumbers,longbibliography,superscriptaddress]{revtex4-2}
\usepackage[utf8]{inputenc}
\usepackage[english]{babel}				

\usepackage{amsmath}
\usepackage{amssymb}
\usepackage{mathtools}					
\usepackage{xcolor}						

\usepackage[colorlinks=true,urlcolor=blue,citecolor=blue,linkcolor=blue]{hyperref}

\newcommand{\dform}[1]{\boldsymbol{#1}}
\newcommand{\dfom}{\dform{\omega}}
\newcommand{\dfPhi}{\dform{\Phi}}
\newcommand{\dfPsi}{\dform{\Psi}}
\newcommand{\dfB}{\dform{B}}
\newcommand{\dfC}{\dform{C}}
\newcommand{\dfR}{\dform{R}}
\newcommand{\dfW}{\dform{W}}
\newcommand{\dfX}{\dform{X}}
\newcommand{\dfY}{\dform{Y}}
\newcommand{\dfZ}{\dform{Z}}

\newcommand{\oddvar}[1]{\overline{#1}{}}

\newcommand\dex{\mathrm{d}}            
\newcommand\dint[1]{#1\lrcorner}       
\newcommand{\vfre}{\boldsymbol{e}}     
\newcommand{\cofr}{\dform{\vartheta}}  
\newcommand{\LCten}{\mathcal{E}}       

\newcommand{\irrW}[1]  {{}^{\scriptscriptstyle(#1)\!}W{}}
\newcommand{\irrZ}[1]  {{}^{\scriptscriptstyle(#1)\!}Z{}}
\newcommand{\irrdfW}[1]{{}^{\scriptscriptstyle(#1)\!}\dfW}
\newcommand{\irrdfZ}[1]{{}^{\scriptscriptstyle(#1)\!}\dfZ}
 
\newcommand\Ri{R}
\newcommand\cR{\tilde{R}}
\newcommand\hR{\hat{R}}
\newcommand\psR{\oddvar{R}}
\newcommand\wR{X}
\newcommand\zR{Y}
 
\begin{document}

\title{On parity-odd sector in metric-affine theories}

\author{Jose Beltr\'an Jim\'enez}
\email{jose.beltran@usal.es}
\affiliation{Departamento de F\'isica Fundamental and IUFFyM, Universidad de Salamanca, E-37008 Salamanca, Spain.}

\author{Alejandro Jim\'enez-Cano}
\email{alejandro.jimenez.cano@ut.ee}
\affiliation{Laboratory of Theoretical Physics, Institute of Physics, University of Tartu, W. Ostwaldi 1, 50411 Tartu, Estonia.}

\author{Yuri N. Obukhov}
\email{obukhov@ibrae.ac.ru}
\affiliation{Russian Academy of Sciences, Nuclear Safety Institute, B.Tulskaya 52, 115191 Moscow, Russia.}

\date{\today}

\begin{abstract}
We undertake the construction of quadratic parity-violating terms involving the curvature in the four-dimensional metric-affine gravity. We demonstrate that there are only 12 linearly independent scalars, plus an additional one that can be removed by using the Pontryagin invariant. Several convenient bases for this sector are provided in both components and differential form notation. We also particularize our general findings to some constrained geometries like Weyl-Cartan and metric-compatible connections.
\end{abstract}

\maketitle

\section{Introduction}

General Relativity (GR) provides the standard geometrical description of the gravitational interaction by means of the spacetime metric. A distinctive property of GR that conforms one of its very foundational aspects is the equivalence principle, which has recently been tested to an astonishing precision of $10^{-15}$ by the MICROSCOPE collaboration \cite{MICROSCOPE:2022doy}. In accordance with this principle, gravity cannot distinguish between different types of matter, which thus manifests its universal character. It is precisely this alluring property what invites to develop a geometrical framework for gravity. In this context, frames become an important aspect of the gravitational interaction to the extent that they can be promoted to the status of fundamental objects. Furthermore, the frames are endowed with an internal Lorentz invariance that naturally introduces a Minkowski structure in the tangent space. Equipped with this set-up, it is then natural to use the field-theoretic machinery to construct gravity as a gauge theory. Since frames are associated to general linear transformations, the relevant group to consider is ${\rm GL}(4,\mathbb{R})$ that leads to the metric-affine gravity (MAG) framework (see e.g. \cite{Hehl1995}). In this approach, the corresponding gauge field strength of the connection is identified with the curvature, in terms of which we can build a quadratic Lagrangian, as in Yang-Mills theory. Such curvature square Lagrangians have been extensively studied in MAG, both in the full metric-affine arena as well as in some restricted frameworks. In most cases, however, emphasis is given to the parity-preserving sector. Currently, there is a considerable growing interest in the parity violation effects in the high-energy physics and in the gravity theory. The early discussions of this issue go back to the works of Kobzarev and Okun \cite{KobzarevOkun} and Leitner and Okubo \cite{Leitner:1964,Haridass:1977,Peres:1962,Peres:1978} in relation with the search of a possible nontrivial gravitoelectric dipole moment of elementary particles, see Mashhoon \cite{Mashhoon:2000} for a review. In the context of gauge gravity,  cosmological applications have recently been explored for the parity-odd scalar mode within the spectrum of Poincaré gravity \cite{Dombriz2021}. However, the previous results were obtained from an action containing only parity-even invariants. In the gauge gravity approach, the parity-violating modifications of gravitational models were considered in \cite{Purcell:1978, Hojman:1980, Hoissen:1983, Mukhopadhyaya:2002}. It is worthwhile to mention the highly interesting manifestations of the parity violation effects that are predicted for gravitational waves \cite{Alexander:2018,Conroy:2019,Nishizawa:2018,Qiao:2019,Yoshida:2018,Zhao:2020}. 

In view of the growing interest in parity-violating effects from the gravity sector, here we undertake a first step towards the construction of general 
 extensions for the metric-affine gravity theory \cite{Hehl1995}. Following the philosophy explained above of building Yang-Mills type of Lagrangians, we will focus on the curvature square invariants, thereby generalizing the recent discussion \cite{Obukhov:2021} and \cite{Iosifidis:2021,Iosifidis:2022a,Iosifidis:2022b} of the torsion and nonmetricity square and linear curvature models of the Hilbert-Einstein type. 

The structure of the paper: In Sec.~\ref{irr}, we analyze the irreducible decomposition of curvature tensor, thus deriving the basic building blocks from which in Sec.~\ref{cur} we construct the complete set of the parity-odd curvature square invariants. Some special geometries are considered in Sec.~\ref{spec}. In Sec.~\ref{concl}, we discuss our results and collect the conclusions. Finally, in Appendix \ref{app:proof}, we prove some useful relations.

Our basic notation and conventions are as follows. We will use the mostly minus signature for the spacetime metric $(+,-,-,-)$ and the natural units  $c=1$. We also introduce the notation $H_{(\mu\nu)}:= \frac{1}{2!}(H_{\mu\nu}+H_{\nu\mu})$ and $H_{[\mu\nu]}:= \frac{1}{2!}(H_{\mu\nu}-H_{\nu\mu})$, and analogously for an object with $n$ indices ($2!\to n!$). The letters of the Greek alphabet label coordinate (i.e., holonomic) spacetime indices, whereas the Latin alphabet is used for the anholonomic frame components. The covariant derivative and the curvature tensor are defined by \eqref{eq:nabla} and \eqref{eq:defR}, respectively. For the Hodge dual of a $k$-form $\dfB=\frac{1}{k!}B_{a_1...a_k} \cofr^{a_1}\wedge...\wedge\cofr^{a_k}$, we use $\star \dfB = \frac{1}{k!(D-k)!} B_{b_1...b_k} \LCten^{b_1...b_k}{}_{a_1...a_{D-k}} \cofr^{a_1}\wedge...\wedge\cofr^{a_{D-k}}$,
where $\LCten_{c_1...c_D}$ is the totally antisymmetric Levi-Civita tensor and $D$ is the dimension of the manifold (in our case $D=4$). The parity-odd objects are marked with an overline: $\psR$, $\oddvar{\phi}_{ab}$, e.g.

\section{Irreducible decomposition of the curvature}\label{irr}

In this section we will introduce the main building blocks that we will employ in our construction of parity-odd invariants. We will commence by reminding some basic aspects of the metric-affine framework and, then, the irreducible decomposition of the curvature will be provided. For completeness, we will do it in both, tensor and exterior calculus formalisms. We will closely follow the treatment in \cite{Hehl1995}.

\subsection{The curvature tensor and the curvature 2-form}

Let us consider an arbitrary smooth 4-dimensional manifold $(M, g, \Gamma)$ equipped with a metric $g_{\mu\nu}$ and an arbitrary  connection $\Gamma_{\mu\nu}{}^\rho$. This connection defines the parallel transport of tensors on $M$, as well as the corresponding covariant derivative that acts on a mixed tensor $H^\mu{}_\nu$ as follows:
\begin{equation}
    \nabla_\mu H^\nu{}_\rho = \partial_\mu H^\nu{}_\rho + \Gamma_{\mu\sigma}{}^\nu H^\sigma{}_\rho - \Gamma_{\mu\rho}{}^{\sigma} H^\nu{}_\sigma\,. \label{eq:nabla}
\end{equation}
The curvature and the torsion tensors associated to this connection are defined as
\begin{align}
    R_{\mu\nu\rho}{}^\lambda &:= 2 \partial_{[\mu}\Gamma_{\nu]\rho}{}^\lambda + 2 \Gamma_{[\mu|\sigma}{}^\lambda\Gamma_{|\nu]\rho}{}^\sigma \,,\label{eq:defR}\\
    T_{\mu\nu}{}^\rho  &:= 2 \Gamma_{[\mu\nu]}{}^\rho\,,
\end{align}
and they arise from the commutator of covariant derivatives
\begin{equation}
\big[\nabla_\mu, \nabla_\nu\big]H^\rho{}_\lambda = R_{\mu\nu\sigma}{}^\rho H^\sigma{}_\lambda -R_{\mu\nu\lambda}{}^\sigma H^\rho{}_\sigma - T_{\mu\nu}{}^\sigma \nabla_\sigma H^\rho{}_\lambda\,.
\end{equation}
Applying this identity to the case of the metric tensor, one finds that the symmetric part of the curvature tensor is given by
\begin{equation}\label{Rsym}
R_{\mu\nu(\rho\lambda)} = \nabla_{[\mu}Q_{\nu]\rho\lambda} + {\frac 12}\,T_{\mu\nu}{}^\sigma Q_{\sigma\rho\lambda},
\end{equation}
so it is nontrivial only when the nonmetricity $Q_{\sigma\alpha\beta} := -\,\nabla_\sigma g_{\alpha\beta}$ does not vanish. 

Let us now switch to an arbitrary (non-necessarily orthogonal) frame and its corresponding coframe:
\begin{equation}
    \vfre_a = e^\mu{}_a \partial_\mu\,,\qquad \cofr^a = e_\mu{}^a \dex x^\mu\,,
\end{equation}
which fulfill $\dint{\vfre_a}\cofr^b =e^\mu{}_a e_\mu{}^b = \delta^b_a $, where $\dint{}$ is the interior product. The anholonomic components of the metric and the connection can be obtained from the coordinate ones as follows:
\begin{align}
    g_{ab} &= e^\mu{}_a e^\nu{}_b g_{\mu\nu}\,,\\ \label{omGam}
    \omega_{\mu a}{}^b&= \Gamma_{\mu\nu}{}^\rho e^\nu{}_a e_\rho{}^b - e^\nu{}_a\partial_\mu e_\nu{}^b\,.
\end{align}
Moreover if we denote the covariant derivative of $\omega_{\mu a}{}^b$ as $D_\mu$, the two derivatives fulfill $D_\mu v^a=e_\nu{}^a\nabla_\mu v^\nu$ for a vector $v^a=e_\nu{}^av^\nu$. In this sense, the second equation (\ref{omGam}) can be interpreted as a map relating the connection $\omega_{\mu a}{}^b$ in the frame bundle with the connection $\Gamma_{\mu\nu}{}^\rho$ in the tangent bundle. The curvature tensor can then be expressed as:
\begin{equation}
   \big(R_{\mu\nu\rho}{}^\lambda e^\rho{}_a e_\lambda{}^b= \big)\quad R_{\mu\nu a}{}^b = 2 \partial_{[\mu}\omega_{\nu]a}{}^b + 2 \omega_{[\mu|c}{}^b \omega_{|\nu]a}{}^c\,,\label{eq:defRab}
\end{equation}
or, in differential form notation, 
\begin{equation}
    \dfR_a{}^b = \dex \dfom_a{}^b + \dfom_c{}^b\wedge \dfom_a{}^c\,,
\end{equation}
where we have introduced the curvature 2-form and the connection 1-form,
\begin{equation}
    \dfR_a{}^b:= \frac{1}{2} R_{\mu\nu a}{}^b \dex x^\mu \wedge \dex x^\nu\,,\qquad \dfom_a{}^b := \omega_{\mu a}{}^b \dex x^\mu\,.
\end{equation}

\subsection{Irreducible decomposition of the curvature tensor}

Having introduced the curvature, we will proceed now to obtain its irreducible decomposition. In this section we will work with all indices referring to an arbitrary frame. We start by lowering the last index of the curvature and splitting it into symmetric and antisymmetric parts:
\begin{equation}
    R_{cdab} = W_{cdab} + Z_{cdab}\,,
\end{equation}
with $W_{cdab}:=R_{cd[ab]}$ and $Z_{cdab}:=R_{cd(ab)}$. At this point, we will introduce the three independent traces of the curvature that can be constructed and that sometimes go under the names of Ricci, co-Ricci and homothetic tensors:
\begin{equation}
    \Ri_{ab} := R_{acb}{}^c\,,\qquad
    \cR_{ab} := R_{ac}{}^c{}_b\,,\qquad
    \hR_{ab} := R_{abc}{}^c\,,
\end{equation}
and also the scalar and the pseudo-scalar that can be constructed with the curvature components,
\begin{equation}
    R := \Ri_{a}{}^a (=-\cR_{a}{}^a)\,,\qquad
    \psR := \LCten^{abcd} R_{abcd}\,.
\end{equation}
Moreover, to compactify the subsequent expressions, it is also convenient to define the following tensors:
\begin{align}
    \wR_{ab} :=& -\frac{1}{2}(\Ri_{ab} - \cR_{ab})
   = W_{cab}{}^c\,,\label{eq:defX}\\
    \zR_{ab} :=& -\frac{1}{2}\left(\Ri_{ab} + \cR_{ab}-\frac{1}{2}\hR_{ab}\right) 
   = \delta_c^d\left(Z_{dab}{}^c-\frac{1}{4}\delta^c_b Z_{dae}{}^e\right)\,,\label{eq:defY}
\end{align}
as well as their symmetric traceless parts:
\begin{align}
    \psi_{ab} &:= \wR_{(ab)} + \frac{1}{4}g_{ab} R \,,\\ 
    \phi_{ab} &:= \zR_{(ab)} \,,
\end{align}
Finally, we introduce the following symmetric traceless parity-odd tensors:
\begin{align}
    \oddvar{\psi}_{ab} &:= \frac{1}{2}\left(W_{cde (a}\LCten^{cde}{}_{b)} - \frac{1}{4}g_{ab} \psR\right) \,,\\ 
    \oddvar{\phi}_{ab} &:= \frac{1}{2}Z_{cde (a}\LCten^{cde}{}_{b)} \label{eq:defauxf}\,.
\end{align}
Notice that $(\psi_{ab},\phi_{ab})$ and $(\bar{\psi}_{ab},\bar{\phi}_{ab})$ have opposite parity transformation properties and we will exploit it later in the construction of parity-odd quadratic scalars.

After introducing all the above useful objects, we now proceed with the irreducible decomposition under the pseudo-orthogonal group (see \cite{McCrea1992, Hehl1995}), following the conventions in \cite{JCObukhov2020}. \\

\noindent{\bf Antisymmetric sector}\\
The antisymmetric component of the curvature can be split into six irreducible pieces:
\begin{equation}
    W_{cdab}=\sum_{I=1}^{6}\irrW{I}{}_{cdab},
\end{equation}
given by
\begin{align}
    \irrW{2}{}_{cdab} 
        & =-\LCten_{cd[a}{}^e\, \oddvar{\psi}_{b]e}\,,\\
    \irrW{3}{}_{cdab} 
        & =R_{[cdab]}=-\frac{1}{24}\LCten_{cdab}\psR\,,\\
    \irrW{4}{}_{cdab} 
        & =2\psi_{ef}\delta_{[a}^{e}g_{b][c}\delta_{d]}^{f}\,,\\
    \irrW{5}{}_{cdab} 
        & =- 2\wR_{[ef]} \delta_{[a}^{e}g_{b][c}\delta_{d]}^{f}\,,\\
    \irrW{6}{}_{cdab} 
        & =\frac{1}{6}g_{c[a}g_{b]d}R\,,\\
    \irrW{1}{}_{cdab} 
        & =W_{cdab}-\sum_{I=2}^{6}\irrW{I}{}_{cdab}\,.
\end{align}
The characteristic properties of these parts can be described as follows:
\begin{itemize}
\item The 2nd piece is the symmetric traceless pseudo-tensor $\oddvar{\psi}_{ab}$ (which has to do with the failure of pair-exchange symmetry for $W_{cdab}$). 
\item The 3rd piece is the totally antisymmetric piece of the curvature (i.e., the pseudo-scalar $\psR$). 
\item The 4th, 5th and 6th pieces basically contain the trace, the antisymmetric and the traceless symmetric parts of $\wR_{ab}$. 
\item Finally, the 1st piece is the remaining traceless piece (the generalization of the Weyl tensor). 
\end{itemize}
In the Riemannian geometry (i.e., for a Levi-Civita connection), we have the usual identities $\wR_{ab}= -R_{ab}=-R_{(ab)}$, $R_{[abcd]}=0$ and $W_{abcd}=R_{abcd}=R_{cdab}$, so that all components vanish, except for $ \irrW{1}{}_{cdab}$,  $\irrW{4}{}_{cdab}$ and  $\irrW{6}{}_{cdab}$.\\

\noindent{\bf Symmetric sector}\\
Let us now move on to analyze the symmetric part. This can be decomposed into five irreducible components:
\begin{equation}
    Z_{cdab}=\sum_{I=1}^{5}\irrZ{I}{}_{cdab},
\end{equation}
which read
\begin{align}
    \irrZ{2}{}_{cdab} 
        & =-\frac{1}{2}\LCten_{cd(a}{}^e\,\oddvar{\phi}_{b)e}\,,\\
    \irrZ{3}{}_{cdab} 
        & =-\frac{1}{3}\zR_{[ef]} (4\delta_{(a}^{e}g_{b)[c}\delta_{d]}^{f}-g_{ab}\delta_{[c}^{e}\delta_{d]}^{f})\,,\\
    \irrZ{4}{}_{cdab} 
        & =- \phi_{ef}\delta_{(a}^{e}g_{b)[c}\delta_{d]}^{f}\,,\\
    \irrZ{5}{}_{cdab} 
        & =\frac{1}{4}g_{ab}\hR{}_{cd}\,,\\
    \irrZ{1}{}_{cdab} 
        & =Z_{cdab}-\sum_{I=2}^{5}\irrZ{I}{}_{cdab}\,.
\end{align}
As for the antisymmetric sector, the irreducible pieces of the symmetric sector admit the following description:
\begin{itemize}
\item The 2nd part contains the symmetric traceless pseudo-tensor $\oddvar{\phi}_{ab}$.
\item  The 3rd and 4th pieces contain respectively the antisymmetric  and symmetric (automatically traceless) parts of $\zR_{ab}$.
    \item the 5th piece contains the homothetic curvature $\hR{}_{ab}$.
    \item Finally, the 1st piece stores the remaining non-trivial part of $Z_{cdab}$
\end{itemize}
All of these pieces vanish for a Levi-Civita connection by virtue of (\ref{Rsym}), i.e., the symmetric sector fully trivialises for Riemannian geometries.

\subsection{Irreducible decomposition of the curvature 2-form}

Now that we have introduced the irreducible decomposition of the curvature tensor, we will proceed to performing the analogous decomposition in the language of differential forms. The two decompositions are of course equivalent, but it is useful to have them both since depending on the specific application one could be more advantageous than the other. First we expand the curvature 2-form into symmetric and antisymmetric parts,
\begin{equation}
    \dfR_{ab} = \dfW_{ab} + \dfZ_{ab}\,,
\end{equation}
with $\dfW_{ab}:=\frac{1}{2}W_{cdab}\cofr^c\wedge\cofr^d$ and $\dfZ_{ab}:=\frac{1}{2}Z_{cdab}\cofr^c\wedge\cofr^d$. 

Introducing the 2-forms $\irrdfW{I}_{ab}:=\frac{1}{2}\irrW{I}_{cdab}\cofr^c\wedge\cofr^d$ and $\irrdfZ{I}_{ab}:=\frac{1}{2}\irrZ{I}_{cdab}\cofr^c\wedge\cofr^d$, we find explicitly:
\begin{align}
  \irrdfW{2}{}_{ab} & = -\star\left(\cofr_{[a}\wedge\oddvar{\dfPsi}{}_{b]}\right)\,,\\
  \irrdfW{3}{}_{ab} & = -\frac{1}{24}\psR\star\cofr_{ab}\,,\\
  \irrdfW{4}{}_{ab} & = -\cofr_{[a}\wedge\dfPsi_{b]}\,,\\
  \irrdfW{5}{}_{ab} & = -\frac{1}{2}\cofr_{[a}\wedge\dint{\vfre_{b]}}(\cofr\dfX)\,,\\
  \irrdfW{6}{}_{ab} & = \frac{1}{12}R \cofr_{ab}\,,\\
  \irrdfW{1}{}_{ab} & = \dfW_{ab} -\sum_{I=2}^{6}\irrdfW{I}{}_{ab}\,,\\
  \irrdfZ{2}{}_{ab} & = -\frac{1}{2}\star\left(\cofr_{(a}\wedge\oddvar{\dfPhi}{}_{b)}\right)\,,\\
  \irrdfZ{3}{}_{ab} & = \frac{1}{6}\left[2\cofr_{(a}\wedge\dint{\vfre_{b)}}(\cofr\dfY)-g_{ab}(\cofr\dfY)\right]\,,\\
  \irrdfZ{4}{}_{ab} & = \frac{1}{2}\cofr_{(a}\wedge\dfPhi{}_{b)}\,,\\
  \irrdfZ{5}{}_{ab} & = \frac{1}{4}g_{ab}\dfZ\,,\\
  \irrdfZ{1}{}_{ab} & = \dfZ_{ab}-\sum_{\mathrm{I}=2}^{5}\irrdfZ{I}{}_{ab}\,,
\end{align}
where we are using the convenient differential forms:
\begin{align}
    &(\cofr\dfX)     := -X_{[ab]}\cofr^a\wedge\cofr^b\,,\qquad
    (\cofr\dfY)     := -Y_{[ab]}\cofr^a\wedge\cofr^b\,,\qquad
    \dfZ:=\dfZ_c{}^c=\frac{1}{2}\hR_{ab}\cofr^a\wedge\cofr^b\,,\nonumber\\
    &\dfPsi_{a}   := \psi_{ab} \cofr^b \,,\qquad
    \dfPhi_{a}   := \phi_{ab}\cofr^b \,,\qquad
    \oddvar{\dfPsi}_{a}:= \oddvar{\psi}_{ab}\cofr^b \,,\qquad
    \oddvar{\dfPhi}_{a}:= \oddvar{\phi}_{ab}\cofr^b \,,
\end{align}
defined in terms of the tensors in \eqref{eq:defX}-\eqref{eq:defauxf} and the homothetic curvature $\hR_{ab}$. These objects coincide with the ones employed in \cite{JCObukhov2020} (see the Appendix of that paper).

\section{The parity-odd curvature-square sector of metric-affine gravity}\label{cur}

Now that we have the irreducible pieces of the curvature, we can proceed to the construction of Lagrangians that are quadratic in the curvature and break parity. The decomposition simplifies the problem because at the quadratic order the sectors of different parity cannot mix (as discussed below) and, thus, they can be treated separately. For practical purposes, it is more suitable to work in terms of objects with a lower number of indices. Of course, one can construct the set of invariants by using the full irreducible components. However, it is more convenient to find a simpler object (with fewer indices) inside each of them containing all of its information, and use these objects instead.

We start by noticing that $\irrW{3}$ and $\irrW{6}$ are the pieces that correspond to the pseudo-scalar and the scalar part of the curvature, respectively, so we can equivalently work with
\begin{align}
    \irrW{3}_{abcd}&\leftrightarrow \psR\,, \\
    \irrW{6}_{abcd}&\leftrightarrow R \label{eq:W6newVar}\,.
\end{align}

Now we focus on the traces of the curvature that give the Ricci, co-Ricci and homothetic tensors defined above $\{\Ri_{ab}, \cR_{ab}, \hR_{ab}\}$. This set of objects is linearly equivalent to $\{\wR_{ab}, \zR_{ab}, \hR_{ab}\}$. Moreover, within these three objects, there is only one independent trace that coincides with the usual Ricci scalar already taken into account in \eqref{eq:W6newVar}. Thus, we only need to take care of the symmetric traceless parts of  $\{\wR_{ab}, \zR_{ab}\}$ and the antisymmetric parts of $\{\wR_{ab}, \zR_{ab}, \hR_{ab}\}$, namely
\begin{equation}
    \{\wR_{[ab]}, \psi_{ab}, \zR_{[ab]}, \phi_{ab}, \hR_{ab }\}\,,
\end{equation}
which relate to the irreducible pieces as follows
\begin{align}
    \irrW{4}_{abcd}&\leftrightarrow \psi_{ab}\,, \\
    \irrW{5}_{abcd}&\leftrightarrow X_{[ab]}\,,\\
    \irrZ{3}_{abcd}&\leftrightarrow \phi_{ab}\,, \\
    \irrZ{4}_{abcd}&\leftrightarrow Y_{[ab]}\,,\\
    \irrZ{5}_{abcd}&\leftrightarrow \hR_{ab}\,.
\end{align}

It only remains to analyze the first two irreducible parts of both sectors, i.e., the set $\{\irrW{1},\irrW{2},\irrZ{1},\irrZ{2}\}$. Together, these four components constitute the totally traceless part of the curvature. Interestingly, two of them are also characterized by symmetric traceless pseudo-tensors with two indices:
\begin{align}
    \irrW{2}_{abcd}&\leftrightarrow \oddvar{\psi}_{ab}\,, \\
    \irrZ{2}_{abcd}&\leftrightarrow \oddvar{\phi}_{ab}\,.
\end{align}
Finally, the remaining components $\{\irrW{1},\irrZ{1},\}$ cannot be expressed in terms of simpler objects.

At this point, it is useful to separate our variables into four sets with different number of indices and symmetries:
\begin{align}
    S_{0}&:=\{R,\ \psR\}, \label{eq:set1}\\
    S_{2}^\text{a}&:=\{X_{[ab]},Y_{[ab]}, \hR_{ab}\},\\
    S_{2}^\text{s}&:=\{\phi_{ab},\psi_{ab}, \oddvar{\phi}_{ab}, \oddvar{\psi}_{ab}\},\\
    S_{4}&:=\{\irrW{1}_{abcd}, \irrZ{1}_{abcd}\}\,.\label{eq:set4}
\end{align}
It is worthwhile to recall that the elements of $S_{2}^\text{a}$ and $S_{2}^\text{s}$ are traceless, and that the elements of $S_{4}$ are completely traceless (i.e. traceless in any pair of indices) and are subject to the constraints
\begin{equation}
    \irrW{1}_{[abcd]}=\irrZ{1}_{[abcd]}=0\,,\qquad \irrW{1}_{[abc]d}=\irrZ{1}_{[abc]d}=0\,,\qquad \irrW{1}_{abcd}=\irrW{1}_{cdab}\,.\label{eq:propW1Z1}
\end{equation}
After a careful look into these properties, the parity of the variables, their symmetries and the allowed index structures, it is not difficult to realize that the parity-odd quadratic combinations between objects belonging to distinct sets \eqref{eq:set1}-\eqref{eq:set4} are all vanishing. Another way to see this is as follows. First, recall that the objects within each of the $S$-subspaces belong to different irreducible representations (irreps) of the pseudo-orthogonal group. The reason why mixings between different $S$'s are zero is that such mixings would involve contractions with the metric and the Levi-Civita tensor (i.e., the invariants of the group in an appropriate frame) that are vanishing for all the irreps. In fact, the two elements of $S_4$ belong to different irreps, so all combination mixing $\irrW{1}$ and $\irrZ{1}$ are vanishing, as we will confirm in Section \ref{eq:invS4}. Having said this, we can start with the study of the quadratic combinations for each of the sets \eqref{eq:set1}-\eqref{eq:set4} separately.

\subsection{Quadratic invariants containing the scalar and the pseudo-scalar}

It is straightforward to conclude that the only possible parity-odd quadratic invariant involving the elements of $S_{0}$ is the product of the scalar and the pseudo-scalar:
\begin{equation}
    R \psR,
\end{equation}
which corresponds to
\begin{equation}
    (\dfW_{ab}\wedge\irrdfW{3}^{ab} = \dfW_{ab}\wedge\irrdfW{6}^{ab}=)\qquad \irrdfW{3}_{ab}\wedge\irrdfW{6}^{ab} \ =\ -\,{\frac{1}{24}}\Ri \psR\ (\star 1)\,.
\end{equation}
Here $\star 1= \sqrt{|g|}\dex^4 x$ is the canonical volume form.

\subsection{Quadratic invariants involving the elements of $S_2^\text{a}$ }

Due to the fact that a product of two tensors in $S_2^{\text{a}}$ has four indices, exactly as the Levi-Civita tensor, this leaves us with just the following possibilities:
\begin{equation}
    \LCten^{abcd} \{ \zR_{ab}\zR_{cd},\ \wR_{ab}\wR_{cd},\  \zR_{ab}\wR_{cd},\  \zR_{ab}\hR_{cd},\  \wR_{ab}\hR_{cd},\  \hR_{ab}\hR_{cd} \}\,.
    \label{eq:Sa}
\end{equation}
Therefore these 6 invariants correspond to the different combinations of $\irrW{5}$, $\irrZ{3}$ and $\irrZ{5}$:
\begin{align}
    (\dfW_{ab}\wedge\irrdfW{5}^{ab}=)\quad
    \irrdfW{5}_{ab}\wedge\irrdfW{5}^{ab}
        &= \frac{1}{2}(\cofr\dfX)\wedge(\cofr\dfX) ,\\
    (\dfZ_{ab}\wedge\irrdfZ{3}^{ab}=)\qquad\irrdfZ{3}_{ab}\wedge\irrdfZ{3}^{ab}
        &=-\frac{1}{3}(\cofr\dfY)\wedge(\cofr\dfY)\,,\\
    (\dfZ_{ab}\wedge\irrdfZ{5}^{ab}=)\qquad\irrdfZ{5}_{ab}\wedge\irrdfZ{5}^{ab}
        &=\frac{1}{4}\dfZ\wedge\dfZ\,,\\
    \dfR_{ab}\wedge\cofr^a\wedge\big(\dint{\vfre_c}\irrdfW{5}{}^{cb}\big)
        &=\frac{1}{2}(\cofr\dfX)\wedge(\cofr\dfX)+\frac{1}{2}(\cofr\dfX)\wedge(\cofr\dfY)-\frac{1}{4}(\cofr\dfX)\wedge\dfZ\,,\\
    \dfR_{ab}\wedge\cofr^a\wedge\big(\dint{\vfre_c}\irrdfZ{3}{}^{cb}\big)
        &=-\frac{1}{2}(\cofr\dfY)\wedge(\cofr\dfY)-\frac{1}{2}(\cofr\dfX)\wedge(\cofr\dfY)+\frac{1}{4}(\cofr\dfY)\wedge\dfZ\,,\\
    \dfR_{ab}\wedge\cofr^a\wedge\big(\dint{\vfre_c}\irrdfZ{5}{}^{cb}\big)
        &=-\frac{1}{4}(\cofr\dfX)\wedge\dfZ - \frac{1}{4}(\cofr\dfY)\wedge\dfZ+\frac{1}{8}\dfZ\wedge\dfZ\,.
\end{align}

Alternatively, the elements of $S_{2}^\text{a}$ can be further decomposed into selfdual and anti-selfdual with nice transformation properties under parity. For instance, if we call $X^\star_{ab}:=\frac{1}{2}  X_{[cd]} \LCten^{cd}{}_{ab}$, then we can introduce the 1-form\footnote{Here we are writing explicitly the antisymmetrization of indices in $X_{[ab]}$ and $Y_{[ab]}$, because the objects $X_{ab}$ and $Y_{ab}$ (defined in \eqref{eq:defX} and \eqref{eq:defY}) also have a symmetric part. The rest of the objects defined in this paragraph, i.e. those with superscripts $\star$ and $\pm$, are automatically antisymmetric by definition.}
\begin{equation}
    X^{\pm}_{ab}:=\frac{1}{\sqrt{2}}\Big(X_{[ab]}\pm {\rm i} X^\star_{ab}\Big),
\end{equation}
which satisfies ${\rm i}X^{\pm \star}_{ab}=\pm X^{\pm}_{ab}$, and similarly for $Y_{[ab]}$ and $\hR_{ab}$. Since we are interested in parity-odd invariants, we need to consider the scalars $\mathcal{S}^{+}_{ab} \mathcal{T}^{-}{}^{ab}$ for $\mathcal{S}_{ab},\mathcal{T}_{ab}\in\{X_{[ab]},Y_{[ab]},\hat{R}_{ab}\}$. These six independent quantities span the same space as \eqref{eq:Sa} in the alternative basis that exploits the duality properties.

\subsection{Quadratic invariants involving the elements of $S_2^\text{s}$ }

As commented above, the elements of this subspace can be separated into two subsets of different parity:
\begin{equation}
    S_2^{\text{s}} = \{\phi_{ab}, \psi_{ab}\}_+\cup \{\oddvar{\phi}_{ab}, \oddvar{\psi}_{ab}\}_-\,.
\end{equation}
Any invariant constructed with two elements of the same subset requires one Levi-Civita tensor, and since our pieces are symmetric, it must be vanishing. In other words, the only possibilities allowed by the index structure are the combinations between elements of different subsets. Then, there are just four options:
\begin{equation}
    \{\phi_{ab}\oddvar{\phi}^{ab}, \phi_{ab}\oddvar{\psi}^{ab}, \psi_{ab}\oddvar{\phi}^{ab}, \psi_{ab}\oddvar{\psi}^{ab}\}\,.
\end{equation}
Such a basis of invariants is linearly equivalent to the following one, in terms of differential forms:
\begin{align}
    (\dfW_{ab}\wedge\irrdfW{4}^{ab}=\dfW_{ab}\wedge\irrdfW{2}^{ab}=)\quad\irrdfW{2}_{ab}\wedge\irrdfW{4}^{ab}
        &=\dfPsi^a\wedge\star\oddvar{\dfPsi}_a\,,\\
    (\dfZ_{ab}\wedge\irrdfZ{4}^{ab}=\dfZ_{ab}\wedge\irrdfZ{2}^{ab}=)\qquad\irrdfZ{2}_{ab}\wedge\irrdfZ{4}^{ab}
        &=-\frac{1}{2}\dfPhi^a\wedge\star\oddvar{\dfPhi}_a\,,\\
    \dfR_{ab}\wedge\cofr_c\wedge\big(\dint{\vfre^{a}}\irrdfZ{2}{}^{cb}\big)
        &=\frac{1}{2}\dfPhi^a\wedge\star\oddvar{\dfPhi}_a-\dfPsi^a\wedge\star\oddvar{\dfPhi}_a\,,\\
    \dfR_{ab}\wedge\cofr^a\wedge\big(\dint{\vfre_c}\irrdfZ{4}{}^{cb}\big)
        &=-\dfPhi^a\wedge\star\oddvar{\dfPhi}_a-\dfPhi^a\wedge\star\oddvar{\dfPsi}_a\,.
\end{align}

\subsection{Quadratic invariants with elements of $S_{4}$ }\label{eq:invS4}

Finally, it remains to analyze the construction of parity-odd scalars with elements of $S_{4}$. Such terms should consist of contractions of two elements of that space and the Levi-Civita tensor. In view of the properties \eqref{eq:propW1Z1}, we can construct the possible structures as
\begin{align}
    L(B,C)&:=\{ 
    B_{ab}{}^{ef}C_{cd{ef}},\
    B^{ef}{}_{ab}C_{{ef}cd},\
    B^{ef}{}_{ab}C_{cd{ef}},\nonumber \\
    &\qquad 
    B_{ab}{}^{ef}C_{c{ef}d},\
    B^{ef}{}_{ab}C_{c{ef}d},\
    C_{ab}{}^{ef}B_{c{ef}d},\
    C^{ef}{}_{ab}B_{c{ef}d},\nonumber\\
    &\qquad 
    B_{a}{}^{ef}{}_{b}C_{c{ef}d} \}\LCten^{abcd}. \label{eq:List0}
\end{align}
Here $B_{abcd}, C_{abcd} \in S_{4}$, and this $L$ is defined so that the redundant and vanishing invariants in it are omitted. Notice that $L(B,C)=L(C,B)$, and hence we only need to check each of the three possible pairs of elements of $S_4$.

The list above can be further reduced, by using the following properties:
\begin{align}
B_{a}{}^{ef}{}_{b}C_{cefd}\LCten^{abcd}
    &\propto B_{ab}{}^{ef}C_{cdef}\LCten^{abcd}\,,\nonumber\\
B_{ab}{}^{ef} C_{cefd}\LCten^{abcd}
    &\propto B_{ab}{}^{ef}C_{cdef}\LCten^{abcd}\,,\nonumber\\
B^{ef}{}_{ab} C_{cefd}\LCten^{abcd}
    &\propto  B^{ef}{}_{ab}C_{cdef}\LCten^{abcd}\,.\label{eq:propBCE}
\end{align}
These can be easily proven with the help of the differential form notation (see Appendix \ref{app:proof} for more details). As a consequence, only the first line of \eqref{eq:List0} contains independent invariants. Therefore,
\begin{equation}
    L(B,C)=\{ 
    B_{ab}{}^{ef}C_{cd{ef}},\
    B^{ef}{}_{ab}C_{{ef}cd},\
    B^{ef}{}_{ab}C_{cd{ef}}\}\LCten^{abcd}. \label{eq:ListReady}
\end{equation}
Notice that due to the symmetry of $\irrZ{1}_{abcd}$ in the last two indices, this expression is showing that $\irrW{1}$ and $\irrZ{1}$ cannot mix up, in agreement with the discussion after \eqref{eq:W6newVar}. This leaves us with just two possibilities:
\begin{itemize}
    \item Case $(B,C)=(\irrW{1},\irrW{1})$. Thanks to the symmetry under exchange of pairs, this reduces to:
    \begin{equation}
        L(\irrW{1},\irrW{1})= \{\irrW{1}_{ab}{}^{ef} \irrW{1}_{cd{ef}} \LCten^{abcd}\}\,.
    \end{equation}
    This combination corresponds in differential form notation to the product
    \begin{equation}
        \irrW{1}^\star_{abcd}\irrW{1}^{abcd} := \frac{1}{2}\irrW{1}_{ab}{}^{ef} 
        \irrW{1}_{cd{ef}}\LCten^{abcd} \qquad \propto \qquad  \irrdfW{1}_{ab}\wedge\irrdfW{1}^{ab}\,.\label{eq:W1W1inv}
    \end{equation}
Notice that the star in the superscript only indicates that the Hodge dual has been calculated with respect to the first two indices.
    \item Case $(B,C)=(\irrZ{1},\irrZ{1})$. The symmetry in the last two indices allows to eliminate the second and the third elements of \eqref{eq:ListReady}, so we arrive at a similar situation as in the previous case:
    \begin{equation}
        L(\irrZ{1},\irrZ{1})= \{\irrZ{1}_{ab}{}^{ef}\irrZ{1}_{cd{ef}} \LCten^{abcd}\}\,.
    \end{equation}
    Thus,
    \begin{equation}
        \irrZ{1}^\star_{abcd} \irrZ{1}^{abcd} := \frac{1}{2}\irrZ{1}_{ab}{}^{ef} 
        \irrZ{1}_{cd{ef}}\LCten^{abcd} \qquad \propto \qquad  \irrdfZ{1}_{ab}\wedge\irrdfZ{1}^{ab}\,.\label{eq:Z1Z1inv}
    \end{equation}
\end{itemize}
Another way of reaching the same result would be to exploit again the fact that $\irrW{1}$ and $\irrZ{1}$ belong to different irreps of the pseudo-orthogonal group so they cannot mix at quadratic order. Since we are seeking to construct parity-odd scalars, then the only possibilities will be $\irrZ{1}\irrZ{1}^\star$ and $\irrW{1}\irrW{1}^\star$, which are the two combinations that we have found with a more direct method.

\subsection{Pontryagin invariant}

In four dimensions, the (Chern-)Pontryagin invariant \cite{Hehl:1991, Obukhov1995, Cherubini:2002gen} is introduced as
\begin{equation}
    \dfR_a{}^b\wedge\dfR_b{}^a = \LCten^{\mu\nu\rho\lambda}R_{\mu\nu a}{}^b R_{\rho\lambda b}{}^a \sqrt{|g|}\dex^4 x = \epsilon^{\mu\nu\rho\lambda}R_{\mu\nu a}{}^b R_{\rho\lambda b}{}^a \dex^4 x\,,\label{eq:Pontry}
\end{equation}
where $\epsilon^{\mu\nu\rho\lambda}$ is just the Levi-Civita symbol (a constant pseudo-tensor density). Notice that the metric and the coframe disappear from the Lagrangian, which thus depends only on the connection in view of \eqref{eq:defRab}. The combination \eqref{eq:Pontry} is a 4-form that can be written as a total derivative for arbitrary connections \cite{Hehl:1991}:
\begin{equation}
    \dfR_a{}^b\wedge\dfR_b{}^a = \dex \left(\dfR_a{}^b\wedge \dfom_b{}^a + \frac{1}{3} \dfom_a{}^b\wedge\dfom_b{}^c\wedge\dfom_c{}^a \right)\,.
\end{equation}
As a consequence, one can ignore this term if we are just interested in the dynamics of the theory. This allows us to drop one of the invariants found in Sec. \ref{eq:invS4}. In the following, we will use this freedom to remove the term $\irrW{1}\irrW{1}^\star$.

\subsection{The quadratic parity-odd metric-affine Lagrangian}

As a result of our previous analysis, the dimension of the vector space of parity-odd curvature-squared invariants is 12 (excluding one that corresponds to the Pontryagin invariant) so this sector will contain 12 independent parameters. One possible choice of the basis is\footnote{
    In case we are interested in keeping the boundary term, one can include for example $\irrW{1}_{ab}{}^{ef}\irrW{1}_{cdef}\LCten^{abcd}$ in \eqref{eq:MAGbasis1}, $R_{ab}{}^{ef}R_{cdfe}\LCten^{abcd}$ in \eqref{eq:MAGbasis2}, $\dfW_{ab}\wedge\dfW^{ab}$ in \eqref{eq:MAGbasis3} or $\irrdfW{1}_{ab}\wedge\irrdfW{1}^{ab}$ in \eqref{eq:MAGbasis4}.}
\begin{align}
    \mathcal{L}_{\text{odd}}=& \alpha R \psR + a_1  X^{ab}X^\star_{ab} +a_2  Y^{ab}X^\star_{ab} +a_3 \hR^{ab} X^\star_{ab} +a_4  Y^{ab}Y^\star_{ab} +a_5  \hR^{ab}Y^\star_{ab} +a_6 \hR^{ab}\hR^\star_{ab} \nonumber\\
    & +s_1 \psi^{ab}\oddvar{\phi}_{ab}+s_2 \phi^{ab}\oddvar{\psi}_{ab}+s_3\oddvar{\phi}^{ab}\phi_{ab}+ s_4\oddvar{\psi}^{ab}\psi_{ab}+\beta\,
    \irrZ{1}^{abcd}\,\irrZ{1}^\star_{abcd}.\label{eq:MAGbasis1}
\end{align}

We can express the general parity-odd quadratic Lagrangian in terms of more familiar objects (Ricci, co-Ricci, homothetic and Riemann tensors) as follows:
\begin{align}
   \mathcal{L}_{\text{odd}}= &\Big(b_1 R R_{abcd}+b_2\Ri_{ab}\Ri_{cd}+b_3 \cR_{ab}\cR_{cd}+b_4  \Ri_{ab}\cR_{cd}+b_5  \Ri_{ab}\hR_{cd}+b_6 \cR_{ab}\hR_{cd}+b_7  \hR_{ab}\hR_{cd}\nonumber\\
    &\quad +b_8R_{abc}{}^e\Ri_{(de)}+ b_9R_{abc}{}^e\cR_{(de)}+ b_{10}R_{ab}{}^e{}_c\Ri_{(de)}+b_{11} R_{ab}{}^e{}_c\cR_{(de)}+ 
    b_{12} R_{ab}{}^{ef} 
    R_{cdef}\Big)\LCten^{abcd}\,.\label{eq:MAGbasis2}
\end{align}

We can alternatively express the parity-odd Lagrangian in the language of differential forms\footnote{The Lagrangian density and its corresponding Lagrangian form are related via ${\bf L}=\mathcal{L} \sqrt{|g|} \dex^4x$.}
\begin{align}
    {\bf L}_{\rm odd}=&\quad
   c_1 \dfZ_{ab}\wedge\dfZ^{ab}+
    c_2\irrdfW{3}_{ab}\wedge\irrdfW{6}^{ab}+ c_3\dfPhi^a\wedge\star\oddvar{\dfPhi}_a+
    c_4\dfPsi^a\wedge\star\oddvar{\dfPsi}_a+
    c_5\dfPsi^a\wedge\star\oddvar{\dfPhi}_a+
    c_6\dfPhi^a\wedge\star\oddvar{\dfPsi}_a\nonumber\\
    &
    +c_7(\cofr\dfX)\wedge(\cofr\dfX)+
    c_8(\cofr\dfX)\wedge(\cofr\dfY)+
    c_9(\cofr\dfY)\wedge(\cofr\dfY)+
    c_{10}(\cofr\dfX)\wedge\dfZ+
    c_{11}(\cofr\dfY)\wedge\dfZ+
    c_{12}\dfZ\wedge\dfZ \,.\label{eq:MAGbasis3}
\end{align}
Let us recall that $\irrdfW{3}_{ab}\wedge\irrdfW{6}^{ab}\propto\Ri\psR\ (\star 1)$. This Lagrangian can equivalently be written as
\begin{align}
    {\bf L}_{\rm odd}=&\quad
   d_1 \irrdfW{2}_{ab}\wedge\irrdfW{4}^{ab}+
    d_2\irrdfW{3}_{ab}\wedge\irrdfW{6}^{ab}+
    d_3\irrdfW{5}_{ab}\wedge\irrdfW{5}^{ab}\nonumber\\
    &+d_4\irrdfZ{1}_{ab}\wedge\irrdfZ{1}^{ab}+
    d_5\irrdfZ{2}_{ab}\wedge\irrdfZ{4}^{ab}+
    d_6\irrdfZ{3}_{ab}\wedge\irrdfZ{3}^{ab}+
    d_7\irrdfZ{5}_{ab}\wedge\irrdfZ{5}^{ab}\nonumber\\
    &+d_8\dfR_{ab}\wedge\cofr^a\wedge\big(\dint{\vfre_c}\irrdfW{5}{}^{cb}\big)+ d_9\dfR_{ab}\wedge\cofr^a\wedge\big(\dint{\vfre_c}\irrdfZ{3}{}^{cb}\big)+
    d_{10}\dfR_{ab}\wedge\cofr^a\wedge\big(\dint{\vfre_c}\irrdfZ{4}{}^{cb}\big)
    \nonumber\\
    &+d_{11}\dfR_{ab}\wedge\cofr^a\wedge\big(\dint{\vfre_c}\irrdfZ{5}{}^{cb}\big)+
    d_{12}\dfR_{ab}\wedge\cofr_c\wedge\big(\dint{\vfre^{a}}\irrdfZ{2}{}^{cb}\big)\,.\label{eq:MAGbasis4}
\end{align}

\section{Special geometries}\label{spec}

In this section we will particularize the general parity-odd Lagrangian obtained above to some particular geometries, thus recovering some known results in the literature.

\subsection{Weyl-Cartan gravity}

We will first particularize to the Weyl-Cartan geometries that are characterized by a nonmetricity tensor that is fully determined by a vector field $V_\mu$ so that $\nabla_\mu g_{\nu\rho}=V_\mu g_{\nu\rho}$. In this geometry, the traceless part of $Z_{cdab}$ vanishes and only the irreducible piece $\irrZ{5}$ survives. Moreover, the co-Ricci tensor is related to the Ricci tensor as $\cR_{ab} = -\Ri_{ab}$ in this geometry. All of these properties imply the following vanishing pieces
\begin{equation}         
    \irrZ{1}_{cdab}=\zR_{ab}=\phi_{ab}=\oddvar{\phi}_{ab}=0\,,
\end{equation}
namely,
\begin{equation}          
    \irrdfZ{1}_{ab}=(\cofr\dfY)=\dfPhi_{a}=\oddvar{\dfPhi}_{a}=0\,,
\end{equation}

Then, one can  write down the analogue of \eqref{eq:MAGbasis1}-\eqref{eq:MAGbasis4} in the Weyl-Cartan case
\begin{align}
    \mathcal{L}^{\rm WC}_{\text{odd}}=\;& \alpha R \psR+ a_1 X^{ab}X^\star_{ab}+a_3 X^{ab} \hR^\star_{ab} + a_6 \hR^{ab}\hR^\star_{ab} +s_4\oddvar{\psi}^{ab}\psi_{ab}
    \\
    =\;&\Big(b_1 R R_{abcd}+b_2\Ri_{ab}\Ri_{cd}+b_5  \Ri_{ab}\hR_{cd}+b_7  \hR_{ab}\hR_{cd}
    +b_8R_{abc}{}^e\Ri_{(de)}\Big)\LCten^{abcd}
    ,\label{eq:WCbasis1}
\end{align}
and 
\begin{align}
    {\bf L}_{\rm odd}^{\rm WC}=&\;
    c_2\irrdfW{3}_{ab}\wedge\irrdfW{6}^{ab}+ 
    c_4\dfPsi^a\wedge\star\oddvar{\dfPsi}_a
    +c_7(\cofr\dfX)\wedge(\cofr\dfX)
    +c_{10}(\cofr\dfX)\wedge\dfZ+
    c_{12}\dfZ\wedge\dfZ \,\\
    =&\;d_1 \irrdfW{2}_{ab}\wedge\irrdfW{4}^{ab}+
    d_2\irrdfW{3}_{ab}\wedge\irrdfW{6}^{ab}+
    d_3\irrdfW{5}_{ab}\wedge\irrdfW{5}^{ab}+d_7\irrdfZ{5}_{ab}\wedge\irrdfZ{5}^{ab}+d_{11}\dfR_{ab}\wedge\cofr^a\wedge\big(\dint{\vfre_c}\irrdfZ{5}{}^{cb}\big)
    .\label{eq:WCbasis3}
\end{align}

\subsection{Poincar\'e gravity}

We conclude with the Poincar\'e gauge gravity case in which the connection is metric-compatible: $\nabla_\mu g_{\nu\rho} = 0$. The basis of invariants has already been presented in the literature (see e.g. \cite{Obukhov:1989}), but we also include it for completeness.  In this case, also the trace of $Z_{cdab}$ in the last two (the homothetic curvature, $\hR_{ab}$) vanishes. Thus, we also have to impose $\dfZ=0$. As a result, only three invariants are allowed. The resulting parity-odd Lagrangians are:
\begin{align}
     \mathcal{L}^{\rm PG}_{\text{odd}}=\;& \alpha R \psR+a_1 X^{ab}X^\star_{ab} +s_4\oddvar{\psi}^{ab}\psi_{ab}
    \\
    =\;&\Big(b_1 R R_{abcd}+b_2\Ri_{ab}\Ri_{cd}
    +b_8R_{abc}{}^e\Ri_{(de)}\Big)\LCten^{abcd}
    ,\label{eq:PGbasis1}
\end{align}
and 
\begin{align}
    {\bf L}_{\rm odd}^{\rm PG}=&\;
    c_2\irrdfW{3}_{ab}\wedge\irrdfW{6}^{ab}+ 
    c_4\dfPsi^a\wedge\star\oddvar{\dfPsi}_a
    +c_7(\cofr\dfX)\wedge(\cofr\dfX)\\
    =&\;d_1 \irrdfW{2}_{ab}\wedge\irrdfW{4}^{ab}+
    d_2\irrdfW{3}_{ab}\wedge\irrdfW{6}^{ab}+
    d_3\irrdfW{5}_{ab}\wedge\irrdfW{5}^{ab}
    .\label{eq:PGbasis3}
\end{align}

\section{Discussion} \label{concl}

In this paper we have undertaken the construction of the general Lagrangian containing parity-odd terms that are quadratic in the curvature in four spacetime dimensions. We have analyzed the construction of all the independent parity-odd scalars by resorting to the irreducible components of the curvature under the pseudo-orthogonal group and we have done in both world tensor components and anholonomic differential form languages. We have shown that there are 13 independent terms, but the existence of the Pontryagin invariant in 4 dimensions allows to remove one of them. Thus, our final Lagrangian consists of 12 independent terms that is our main result. We have recovered known results in the literature by particularizing to Weyl-Cartan geometries and Poincar\'e gauge theory.

Our results will be important in the context of a general development of theories of gravity beyond the Einstein framework, including the discussion of fundamental issues such as the equivalence principle \cite{Capozziello:2022}. In a more specialized sense, the results obtained will contribute to the studies of parity-violating effects in cosmology and black hole physics since the obtained Lagrangians complete the quadratic parity-odd sector of the general metric-affine framework. On the other hand, before applications of our Lagrangian can be robustly and reliably obtained, a stability analysis should be performed, since a common problem of quadratic curvature theories in the metric-affine realm is the presence of ghost-like instabilities (either in the linear or in the full non-linear spectrum; see e.g. \cite{JCMaldo2022}).

\begin{acknowledgments}
AJC was supported by the European Regional Development Fund through the Center of Excellence TK133 ``The Dark Side of the Universe'' and by the Mobilitas Pluss postdoctoral grant MOBJD1035. J.B.J. was supported by the Project PGC2018-096038-B-100 funded by the Spanish "Ministerio de Ciencia e Innovaci\'on".
\end{acknowledgments}

\appendix

\section{Proof of \eqref{eq:propBCE}}\label{app:proof}

If we call $\dfB_{ab}=\frac{1}{2}B_{cdab}\cofr^a\wedge\cofr^b$ and $\dfC_{ab}=\frac{1}{2}C_{cdab}\cofr^a\wedge\cofr^b$, we can derive the following proportionality relations:
\begin{align}
B_{a}{}^{ef}{}_{b}C_{cefd}\LCten^{abcd}
    &\propto\Big[(\dint{\vfre^e}\dfB^f{}_b)\wedge\cofr^b\Big]\wedge\Big[(\dint{\vfre_e}\dfC_{fd})\wedge\cofr^d\Big]\nonumber\\
    &\propto -\dfB^f{}_b\wedge(\dint{\vfre^e}\cofr^b)\wedge(\dint{\vfre_e}\dfC_{fd})\wedge\cofr^d-\dfB^f{}_b\wedge\cofr^b\wedge(\dint{\vfre_e}\dfC_{fd})\wedge(\dint{\vfre^e}\cofr^d)\nonumber\\
    &\propto -\dfB^f{}_b\wedge(\dint{\vfre^b}\dfC_{fd})\wedge\cofr^d-0\nonumber\\
    &\propto \dfB^f{}_b\wedge\dfC_{fd}\wedge(\dint{\vfre^b}\cofr^d)\nonumber\\
    &\propto \dfB^{ef}\wedge\dfC_{ef}\qquad \propto B_{ab}{}^{ef}C_{cdef}\LCten^{abcd}\,,\\[1em]
B_{ab}{}^{ef} C_{cefd}\LCten^{abcd}
    &\propto\dfB^{ef}\wedge\Big[(\dint{\vfre_e}\dfC_{fd})\wedge\cofr^d\Big]\nonumber\\
    &\propto
    \dfB^{ef}\wedge\dfC_{fd}\wedge(\dint{\vfre_e}\cofr^d) \nonumber\\
    &\propto \dfB^{df}\wedge \dfC_{fd} \qquad\propto B_{ab}{}^{ef}C_{cdef}\LCten^{abcd}\,,\\[1em]
B^{ef}{}_{ab} C_{cefd}\LCten^{abcd}
    &\propto\Big[(\dint{\vfre^{e}}\dint{\vfre^{f}}\dfB_{ab})\wedge\cofr^a\wedge\cofr^b\Big]\wedge\Big[(\dint{\vfre_e}\dfC_{fd})\wedge\cofr^d\Big]\nonumber\\
    &\propto (...)\wedge(\dint{\vfre^{e}}\dint{\vfre^{f}}\dfB_{eb}) + (...)\wedge(\dint{\vfre^{e}}\dint{\vfre^{f}}\dfB_{ae})\nonumber\\
    &\qquad+ \Big[(\dint{\vfre^{e}}\dint{\vfre^{f}}\dfB_{ab})\wedge\cofr^a\wedge\cofr^b\Big]\wedge\Big[\dfC_{fd}\wedge(\dint{\vfre_e}\cofr^d)\Big] \nonumber\\
    &\propto \Big[(\dint{\vfre^{e}}\dint{\vfre^{f}}\dfB_{ab})\wedge\cofr^a\wedge\cofr^b\Big]\wedge\dfC_{fe} \qquad\propto B^{ef}{}_{ab}C_{cdef}\LCten^{abcd}\,,
\end{align}
where we have used the traceless conditions $\dint{\vfre^a}\dfB_{fa}=0=\dint{\vfre^a}\dfC_{fa}$, the duality $\dint{\vfre_a}\cofr^b=\delta^b_a$, the antisymmetry of the interior product $\dint{\vfre^a}\dint{\vfre_a}=0$, the property $\star (\cofr^a\wedge \cofr^b\wedge \cofr^c\wedge \cofr^d)= \LCten^{abcd}$ and the fact that 5-forms are vanishing in four dimensions. The symbol (...) denotes some combination of forms that we have omitted because it is irrelevant for the proof.

\bibliography{references.bib}

\end{document}